\input amstex
\magnification=1200
\documentstyle{amsppt}
\leftheadtext{}
\rightheadtext{}
\NoBlackBoxes
\define\End{\operatorname{End}}

\define\Hom{\operatorname{Hom}}

\define\Pn{\hat P_t}
\define\Q{\hat Q_t}
\define\R{\hat R_t}
\define\A{\hat A_t}
\define\Bn{\hat B_t}
\define\C{\hat C_t}
\define\spn{\operatorname{span}}

\topmatter
\title On the dynamics of noncanonically coupled
oscillators and its hidden superstructure\endtitle
\author Denis V. Juriev
\endauthor
\affil\eightpoint{\it
Erwin Schr\"odinger Institute f\"ur Mathematische Physik,\linebreak
Pasteurgasse 6/7, Wien, A-1090, \"Osterreich
(Austria)}
\endaffil
\abstract
The classical and quantum dynamics of the noncanonically coupled oscillators
is considered. It is shown that though the classical dynamics is well--defined
for both harmonic and anharmonic oscillators, the quantum one is well--defined
in the harmonic case, admits a hidden (super)Hamiltonian formulation, and thus,
preserves the initial operator relations, whereas a na\"\i ve quantization of
the anharmonic case meets with principal difficulties.
\endabstract
\endtopmatter
\document
The classical and quantum dynamics of Hamiltonian systems is often described by
remarkable algebraic structures such as Lie algebras, their nonlinear
generalizations and (quantum) deformations [1]. It seems that not less
important objects govern a behaviour of the interacting Hamiltonian systems
and that they maybe unravelled in a certain way. There exist several forms of
an interaction of Hamiltonian systems: often it has a potential character,
sometimes it is ruled by a deformation of the Poisson brackets; however, one
of the most intriguing, physically important but mathematically less explored
forms is a nonHamiltonian interaction, which can not be described by
deformations of the standard Hamiltonian data (Poisson brackets and
Hamiltonians). Sometimes, such interaction may be realized by a dependence
of the Poisson brackets of one Hamiltonian system on the state of another.
This is just the magnetic--type interaction, which is realised in systems of
charged bodies interacting via Amp\`ere--Lorentz forces. Such interaction is
universal as a certain "classical mechanics" approximation for the most of
the field (e.g. gravitational or nonabelian gauge) or continuum media theories.
For example, the Amp\`ere--Lorentz--type approximation in the general
relativity is sufficient for the quantitative derivations of the Mercury
perihelion's shift, Lenze--Thirring effect, etc. Also the magnetic--type
nonHamiltonian interactions appear in the classical mechanics itself
(hyroscopic systems).

The pair of noncanonically coupled oscillators is one of the simplest and the
most crucial examples of the nonHamiltonian interaction [2]; this Letter is
devoted to an investigation of the related classical and quantum dynamics. It
is shown that though the classical dynamics is well--defined for both harmonic
and anharmonic case, the quantum one is well-defined for noncanonically coupled
harmonic oscillators, admits a hidden (super)Hamiltonian formulation, and
hence, preserves the initial operator relations (cf.[3]), whereas a na\"\i ve
quantization of anharmonic oscillators meets with principal difficulties.

\head\bf 1. Isotopic pair of noncanonically coupled oscillators.\endhead

\subhead 1.1. General algebraic definitions\endsubhead
Let's describe algebraic objects underlying the dynamics, which we
are interested in.

\definition{Definition 1 {\rm [2] (cf.[4])}} The pair $(V_1,V_2)$ of linear
spaces is called {\it an isotopic pair\/} iff there are defined two mappings
$m_1:V_2\otimes\bigwedge^2V_1\mapsto V_1$ and
$m_2:V_1\otimes\bigwedge^2V_2\mapsto V_2$ such that the mappings $(X,Y)\mapsto
[X,Y]_A=m_1(A,X,Y)$ ($X,Y\in V_1$, $A\in V_2$) and $(A,B)\mapsto
[A,B]_X=m_2(X,A,B)$ ($A,B\in V_2$, $X\in V_1$) obey the Jacobi identity for
all values of a subscript parameter (such operations will be called {\it
isocommutators\/} and the subscript parameters will be called {\it isotopic
elements\/}) and are compatible to each other,
i.e. the identities
$$\align
[X,Y]_{[A,B]_Z}=&\tfrac12([[X,Z]_A,Y]_B+[[X,Y]_A,Z]_B+[[Z,Y]_A,X]_B-\\
&[[X,Z]_B,Y]_A-[[X,Y]_B,Z]_A-[[Z,Y]_B,X]_A)\endalign$$
($X,Y,Z\in V_1$,
$A,B\in V_2$ or $X,Y,Z\in V_2$, $A,B\in V_1$) hold.
\enddefinition

This defintion may be considered as a result of an axiomatization of the
following construction: let $\Cal A$ be an associative algebra (f.e. any matrix
one) and $V_1$, $V_2$ be two linear subspaces in it such that $V_1$ is closed
under the isocommutators $(X,Y)\mapsto [X,Y]_A=XAY-YAX$ with isotopic elements
$A$ from $V_2$, whereas $V_2$ is closed under the isocommutators $(A,B)\mapsto
[A,B]_X=AXB-BXA$ with isotopic elements $X$ from $V_1$.

Isotopic pairs are closely related to the (polarized) anti--Lie triple
systems and Lie superalgebras (cf.[4]). Namely,

\definition{Definition 2} The ternary algebra $V$ with product $[xyz]$ is
called {\it an anti--Lie triple system\/} if (1) $[xyz]=[xzy]$, (2)
$[xyz]+[zxy]+[yzx]=0$, (3) $[[xyz]uv]=[[xuv]yz]+[x[yvu]z]+[xy[zuv]]$.
An anti--Lie triple system $V$ is {\it polarized\/} iff $V=V_1\oplus
V_2$ and $[xyz]=0$ for $y,z\in V_1$ or $y,z\in V_2$.
\enddefinition

If $V$ is an anti--Lie triple system let's put $R_{yz}\in\End(V):
R_{yz}x=[xyz]$. The operators $R_{yz}$ are closed under commutators so that
$\frak g_0(V)=\spn(R_{yz}; y,z\in V)$ is a Lie algebra. The space $\frak
g_0(V)\oplus V$ possesses a natural structure of a Lie superalgebra with
the even part $\frak g_0(V)$ and the odd part $V$ [4]. It will be denoted by
$\frak g(V)$. Polarized anti--Lie triple systems $V=V_1\oplus V_2$ produce
polarized Lie superalgebras $\frak g(V)=\frak g_0(V)\oplus(V_1\oplus V_2)$
such that $[V_i,V_i]_+=0$, $[\frak g(V), V_i]_-\subseteq V_i$ (it should be
marked that there is sometimes asserted that $V_2\simeq V_1^*$ as $\frak
g_0(V)$--modules, however, we shall not do it in general).

An arbitrary isotopic pair has a structure of a polarized anti--Lie triple
system (cf.[4]). Namely, one should put $[xyz]=[z,x]_y$ (iff $z$ belongs to
the same space $V_i$ as $x$) and $[y,x]_z$ (iff $y$ belongs to the same space
$V_i$ as $x$).

An illustrative example to the construction of a Lie superalgebra by an
isotopic pair is convenient. {\it Example\/}: let $H_1$ and $H_2$ be two
linear spaces, $(\Hom(H_1,H_2);\mathbreak\Hom(H_2,H_1))$ is an isotopic pair,
the corresponding Lie superalgebra is isomorphic to $\operatorname{\frak
g\frak l}(n|m)$, $n=\dim H_1$, $m=\dim H_2$.

\subhead 1.2. Nonlinear dynamical equations associated with isotopic
pairs\endsubhead
Note that the isocommutators in an isotopic pair $(V_1,V_2)$
define families of compatible Poisson brackets $\{\cdot,\cdot\}_A$ and
$\{\cdot,\cdot\}_X$ ($A\in V_2$, $X\in V_1$) in the spaces $S^{\cdot}(V_1)$
and $S^{\cdot}(V_2)$, respectively. The compatibility means that a
linear combination of any two Poisson brackets is also a Poisson
bracket.

\definition{Definition 3 {\rm (cf.[2])}} Let's consider two elements $\Cal
H_1$ and $\Cal H_2$ ({\it "Hamiltonians"\/}) in $S^{\cdot}(V_1)$ and
$S^{\cdot}(V_2)$, respectively. The equations
$$\dot X_t=\{\Cal H_1,X_t\}_{A_t},\qquad \dot A_t=\{\Cal H_2,A_t\}_{X_t},$$
where $X_t\in V_1$ and $A_t\in V_2$ are called {\it the (nonlinear) dynamical
equations associated with the isotopic pair $(V_1,V_2)$ and "Hamiltonians"
$\Cal H_1$ and $\Cal H_2$\/}.
\enddefinition

\subhead 1.3. Isotopic pair of noncanonically coupled oscillators\endsubhead
Let's now consider the isotopic pairs of noncanonically coupled oscillators
[2,5]. The space $V_1$ is spanned by the elements $p$, $q$ and $r$ and the
space $V_2$ is spanned by the elements $a$, $b$ and $c$. The isocommutators
have the form

\

\centerline{
$\aligned
[p,q]_a&=2\varepsilon_1 q\\
[p,r]_a&=\varepsilon_2 r\\
[q,r]_a&=0
\endaligned$
$\quad$
$\aligned
[p,q]_b&=2\varepsilon_1 p\\
[p,r]_b&=0\\
[q,r]_b&=-\varepsilon_2 r
\endaligned$
$\quad$
$\aligned
[p,q]_c=\varepsilon_3 r\\
[p,r]_c=0\\
[q,r]_c=0
\endaligned$}

\

\

\centerline{
$\aligned
[a,b]_p&=2\tilde\varepsilon_1 b\\
[a,c]_p&=\tilde\varepsilon_2 c\\
[b,c]_p&=0
\endaligned$
$\quad$
$\aligned
[a,b]_q&=2\tilde\varepsilon_1 a\\
[a,c]_q&=0\\
[b,c]_q&=-\tilde\varepsilon_2 c
\endaligned$
$\quad$
$\aligned
[a,b]_r=\tilde\varepsilon_3 c\\
[b,c]_r=0\\
[a,c]_r=0
\endaligned$}

\

where $\varepsilon_1+\tilde\varepsilon_1=0$,
$\varepsilon_2-\tilde\varepsilon_2=\varepsilon_1-\tilde\varepsilon_1$,
$\varepsilon_3\tilde\varepsilon_3-\varepsilon_2\tilde\varepsilon_2=0$.

The corresponding Lie algebra $\frak g_0(V_1\oplus V_2)$ is spanned (for
generic $\varepsilon_i$, $\tilde\varepsilon_i$) by 6 operators
$R_{p,a}$, $R_{p,b}$, $R_{q,a}$, $R_{q,b}$,
$R_{r,b}=\frac{\varepsilon_2}{\varepsilon_3}R_{p,c}$,
$R_{r,a}=\frac{\varepsilon_2}{\varepsilon_3}R_{q,c}$,
which have the form
$$\align
R_{p,a}=\left(\matrix 2\varepsilon_1 & 0 & 0 \\ 0 & 0 & 0 \\ 0 & 0 &
\varepsilon_2\endmatrix\right),& \quad
R_{p,b}=\left(\matrix 0 & 0 & 0 \\ 2\varepsilon_1 & 0 & 0 \\ 0 & 0 &
0\endmatrix\right), \quad
R_{q,a}=\left(\matrix 0 & -2\varepsilon_1 & 0 \\ 0 & 0 & 0 \\ 0 & 0 &
0\endmatrix\right),\\
R_{q,b}=\left(\matrix 0 & 0 & 0 \\ 0 & -2\varepsilon_1 & 0 \\ 0 & 0 &
-\varepsilon_2\endmatrix\right),& \quad
R_{p,c}=\left(\matrix 0 & 0 & 0 \\ 0 & 0 & 0 \\ \varepsilon_3 & 0 &
0\endmatrix\right), \quad
R_{q,c}=\left(\matrix 0 & 0 & 0 \\ 0 & 0 & 0 \\ 0 & -\varepsilon_3 &
0\endmatrix\right)
\endalign $$
in the basis $(q,p,r)$ and the form
$$\align
R_{p,a}=\left(\matrix 0 & 0 & 0 \\ 0 & 2\tilde\varepsilon_1 & 0 \\ 0 & 0 &
\tilde\varepsilon_2\endmatrix\right),& \quad
R_{p,b}=\left(\matrix 0 & 0 & 0 \\ -2\tilde\varepsilon_1 & 0 & 0 \\ 0 & 0 &
0\endmatrix\right), \quad
R_{q,a}=\left(\matrix 0 & 2\tilde\varepsilon_1 & 0 \\ 0 & 0 & 0 \\ 0 & 0 &
0\endmatrix\right),\\
R_{q,b}=\left(\matrix -2\tilde\varepsilon_1 & 0 & 0 \\ 0 & 0 & 0 \\ 0 & 0 &
-\tilde\varepsilon_2\endmatrix\right),& \quad
R_{p,c}=\left(\matrix 0 & 0 & 0 \\ 0 & 0 & 0 \\ -\tilde\varepsilon_2 & 0 &
0\endmatrix\right), \quad
R_{q,c}=\left(\matrix 0 & 0 & 0 \\ 0 & 0 & 0 \\ 0 & \tilde\varepsilon_2 &
0\endmatrix\right)
\endalign $$
in the basis $(a,b,c)$.

The Lie superalgebra $\frak g(V_1\oplus V_2)$ has a (super)dimension $(6|6)$
and is generated by $R_{p,a}$, $R_{p,b}$, $R_{q,a}$, $R_{q,b}$, $R_{p,c}$,
$R_{q,c}$, $p$, $q$, $r$, $a$, $b$, $c$ with (super)commutation relations
$$\align
[q,p]_+=[q,r]_+=[p,r]_+=[a,b]_+&=[a,c]_+=[b,c]_+=[r,c]_+=0,\\
[p,a]_+=R_{p,a},\ [q,a]_+&=R_{q,a},\ [p,b]_+=R_{p,b},\\ [q,b]_+=R_{q,b},\
[p,c]_+&=R_{p,c},\ [q,c]_+=R_{q,c},\\
[r,a]_+=\tfrac{\varepsilon_2}{\varepsilon_3}R_{q,c},&\
[r,b]_+=\tfrac{\varepsilon_2}{\varepsilon_3}R_{p,c};\\
&\\ \allowdisplaybreak
[R_{p,a},q]_-=2\varepsilon_1q,\ [R_{p,a},p]_-&=0,\
[R_{p,a},r]_-=\varepsilon_2r,\\
[R_{q,a},q]_-=0,\ [R_{q,a},p]_-&=-2\varepsilon_1q,\ [R_{q,a},r]_-=0,\\
[R_{p,b},q]_-=2\varepsilon_1p,\ [R_{p,b},p]_-&=0,\ [R_{p,b},r]_-=0,\\
[R_{q,b},q]_-=0,\ [R_{q,b},p]_-&=-2\varepsilon_1p,\
[R_{q,b},r]_-=-\varepsilon_2r,\\
[R_{p,c},q]_-=\varepsilon_3r,\ [R_{p,c},p]_-&=0,\ [R_{p,c},r]_-=0,\\
[R_{q,c},q]_-=0,\ [R_{q,c},p]_-&=-\varepsilon_3r,\ [R_{q,c},r]_-=0,\\
[R_{p,a},a]_-=0,\ [R_{p,a},b]_-&=2\tilde\varepsilon_1b,\
[R_{p,a},c]_-=\tilde\varepsilon_2c,\\
[R_{q,a},a]_-=0,\ [R_{q,a},b]_-&=2\tilde\varepsilon_1a,\ [R_{q,a},c]_-=0,\\
[R_{p,b},a]_-=-2\tilde\varepsilon_1b,\ [R_{p,b},b]_-&=0,\ [R_{p,b},c]_-=0,\\
[R_{q,b},a]_-=-2\tilde\varepsilon_1a,\ [R_{q,b},b]_-&=0,\
[R_{q,b},c]_-=-\tilde\varepsilon_2c,\\
[R_{p,c},a]_-=-\tilde\varepsilon_2c,\ [R_{p,c},b]_-&=0,\ [R_{p,c},c]_-=0,\\
[R_{q,c},a]_-=0,\ [R_{q,c},b]_-&=\tilde\varepsilon_2c,\ [R_{q,c},c]_-=0;\\
&\\ \allowdisplaybreak
[R_{p,a},R_{p,b}]_-=-2\varepsilon_1R_{p,b},\
[R_{p,a},R_{q,a}]_-&=2\varepsilon_1R_{q,a},\ [R_{p,a},R_{p,b}]_-=0,\\
[R_{p,a},R_{p,c}]_-=\tilde\varepsilon_2R_{p,c},\
[R_{p,a},R_{q,c}]_-&=\varepsilon_2R_{q,c},\
[R_{p,b},R_{q,a}]_-=2\varepsilon_1(R_{q,b}+R_{p,a}),\\
[R_{p,b},R_{q,b}]_-=2\varepsilon_1R_{p,b},\ [R_{p,b},R_{p,c}]_-&=0,\
[R_{p,b},R_{q,c}]_-=2\varepsilon_1R_{p,c},\\
[R_{q,a},R_{q,b}]_-=-2\varepsilon_1R_{q,a},\
[R_{q,a},R_{p,c}]_-&=-2\varepsilon_1R_{p,c},\
[R_{q,a},R_{q,c}]_-=0,\\
[R_{q,b},R_{p,c}]_-=-\varepsilon_2R_{p,c},\
[R_{q,b},R_{q,c}]_-&=-\tilde\varepsilon_2R_{q,c},\ [R_{p,c},R_{q,c}]_-=0.
\endalign$$

The even part of the Lie superalgebra $\frak g(V_1\oplus V_2)$ is isomorphic
to the semidirect sum of $\operatorname{\frak g\frak l}(2,\Bbb C)$ and $\Bbb
C^2$. On the other hand $\frak g(V_1\oplus V_2)$ may be considered as a
semidirect product of the Lie superalgebra $\operatorname{\frak s\frak
l}(2|1,\Bbb C)$ generated by $R_{p,a}$, $R_{p,b}$, $R_{q,a}$, $R_{q,b}$, $p$,
$q$, $a$, $b$ and the $(2|2)$--dimensional vector superspace $V^{2|2}$
generated by $R_{p,c}$, $R_{q,c}$, $r$, $c$.

\head\bf 2. Classical and quantum dynamics of noncanonically coupled oscillators.
\endhead

\subhead 2.1. Classical dynamics of noncanonically coupled harmonic
oscillators\endsubhead
First of all, let's describe the classical dynamics of noncanonically coupled
harmonic oscillators. The dynamical equations with "Hamiltonians" $\Cal
H_1=P^2+Q^2$ and $\Cal H_2=A^2+B^2$ have the form
$$\left\{\aligned
\dot P=&-4\varepsilon_1(Q^2A+PQB)-2\varepsilon_3RQC\\
\dot Q=&4\varepsilon_1(PQA+P^2B)+2\varepsilon_3RPC\\
\dot R=&2\varepsilon_2(PRA-QRB)\endaligned\right.\quad
\left\{\aligned
\dot A=&-4\tilde\varepsilon_1(B^2P+ABQ)-2\tilde\varepsilon_3CBR\\
\dot B=&4\tilde\varepsilon_1(ABP+A^2Q)+2\tilde\varepsilon_3CAR\\
\dot C=&2\tilde\varepsilon_2(ACP-BCQ)\endaligned\right.$$

Note that "Hamiltonians" $\Cal H_1=\Cal I_1^2$ and $\Cal H_2=\Cal I_2^2$
are integrals of motion here, so it is rather convenient to put $P=\Cal
I_1\cos\varphi$, $Q=\Cal I_1\sin\varphi$, $A=\Cal I_2\cos\psi$, $B=\Cal
I_2\sin\psi$. Then
$$\left\{\aligned
\dot\varphi=&2\varepsilon_3RC+4\varepsilon_1\Cal I_1\Cal
I_2\sin(\varphi+\psi)\\
\dot\psi=&2\tilde\varepsilon_3RC+4\tilde\varepsilon_1\Cal I_1\Cal
I_2\sin(\varphi+\psi)\endaligned\right.\quad
\left\{\aligned
\dot R=&2\varepsilon_2\cos(\varphi+\psi)R\\
\dot C=&2\tilde\varepsilon_2\cos(\varphi+\psi)C
\endaligned\right.$$
Let's introduce $\vartheta=\varphi+\psi$, $\chi=\varepsilon_3\psi-
\tilde\varepsilon_3\varphi$ and mark that $\varepsilon_1+
\tilde\varepsilon_1=0$, then
$$\left\{\aligned
\dot\vartheta=&2(\varepsilon_3+\tilde\varepsilon_3)RC\\
\dot\chi=&-4\varepsilon_1\Cal I_1\Cal I_2(\varepsilon_3-\tilde\varepsilon_3)
\sin\vartheta\endaligned\right.$$
Also
$(RC)^{\cdot}=2(\varepsilon_2+\tilde\varepsilon_2)\cos\vartheta(RC)$,
therefore, $(RC)'_{\vartheta}=\frac{\varepsilon_2+\tilde\varepsilon_2}
{\varepsilon_3+\tilde\varepsilon_3}\cos\vartheta$, and
$RC=\Cal L+\frac{\varepsilon_2+\tilde\varepsilon_2}{\varepsilon_3+\tilde
\varepsilon_3}\Cal I_1\Cal I_2\sin\vartheta$,
whereas
$$\dot\vartheta=2\Cal L(\varepsilon_3+\tilde\varepsilon_3)+2\Cal I_1\Cal I_2
(\varepsilon_2+\tilde\varepsilon_2)\sin\vartheta.$$
Here $\Cal L=RC-\frac{\varepsilon_2+\tilde\varepsilon_2}{\varepsilon_3+\tilde
\varepsilon_3}(QA+PB)$ is an integral of motion.
Note that
$(R^{\tilde\varepsilon_2}C^{-\varepsilon_2})^{\cdot}=0$,
so it is convenient to put
$\Lambda=R^{\frac{\tilde\varepsilon_2}{\varepsilon_2+\tilde\varepsilon_2}}
C^{\frac{\varepsilon_2}{\varepsilon_2+\tilde\varepsilon_2}}.$

Then
$$\left\{\aligned
R=&\Lambda(\Cal L-\frac{\varepsilon_2+\tilde\varepsilon_2}{\varepsilon_3+\tilde
\varepsilon_3}\Cal I_1\Cal I_2\sin\vartheta)^{\frac{\varepsilon_2}
{\varepsilon_2+\tilde\varepsilon_2}}\\
C=&\frac1{\Lambda}(\Cal
L-\frac{\varepsilon_2+\tilde\varepsilon_2}{\varepsilon_3+
\tilde\varepsilon_3}\Cal I_1\Cal I_2\sin\vartheta)^{\frac{\tilde\varepsilon_2}
{\varepsilon_2+\tilde\varepsilon_2}}\endaligned\right.$$
$\Cal I_1$, $\Cal I_2$, $\Cal L$ and $\Lambda$ form a complete set of
integrals of motion for generic values of $\varepsilon_i$,
$\tilde\varepsilon_i$.

Let's also denote
$\xi=(\varepsilon_2+\tilde\varepsilon_2)\chi+
2\varepsilon_1(\varepsilon_3-\tilde\varepsilon_3)\vartheta
=[(\varepsilon_2+\tilde\varepsilon_2)\varepsilon_3+
2\varepsilon_1(\varepsilon_3-\tilde\varepsilon_3)]\psi-[(\varepsilon_2+
\tilde\varepsilon_2)\tilde\varepsilon_3+2\tilde\varepsilon_1(\varepsilon_3-
\tilde\varepsilon_3)]\varphi$, then
$$\xi=4\Cal L(\varepsilon^2_3-\tilde\varepsilon^2_3)\varepsilon_1t+\xi_0.$$

\subhead 2.2. Classical dynamics of noncanonically coupled anharmonic
oscillators\endsubhead
The dynamical equations with anharmonic "Hamiltonians"
$\Cal H_1=P^2+V(Q^2)$
and
$\Cal H_2=A^2+V(B^2)$ have the form
$$\left\{\aligned
\dot P=&-V'(Q^2)(4\varepsilon_1(Q^2A+PQB)+2\varepsilon_3RQC)\\
\dot Q=&4\varepsilon_1(PQA+P^2B)+2\varepsilon_3RPC\\
\dot R=&2\varepsilon_2(PRA-QRBV'(Q^2))
\endaligned\right.$$
$$\left\{\aligned
\dot A=&-V'(B^2)(4\tilde\varepsilon_1(B^2P+ABQ)+2\tilde\varepsilon_3CBR)\\
\dot B=&4\tilde\varepsilon_1(ABP+A^2Q)+2\tilde\varepsilon_3CAR\\
\dot C=&2\tilde\varepsilon_2(ACP-BCQV'(B^2))
\endaligned\right.$$

Let's put $S=RC$, $T=AQ+BP$, then
$$\left\{\aligned\dot S=&2S((\varepsilon_2+\tilde\varepsilon_2)PA-QB
(\varepsilon_2V'(Q^2)+\tilde\varepsilon_2V'(B^2)))\\
\dot T=&2S((\varepsilon_3+\tilde\varepsilon_3)PA-QB
(\varepsilon_3V'(Q^2)+\tilde\varepsilon_3V'(B^2)))\endaligned\right.$$
At the same time
$$\left\{\aligned\dot P=&-QV'(Q^2)(4\varepsilon_1T+2\varepsilon_3S)\\
\dot Q=&4\varepsilon_1PT+2\varepsilon_3PS\endaligned\right.\quad
\left\{\aligned
\dot A=&-BV'(B^2)(4\tilde\varepsilon_1T+2\tilde\varepsilon_3S)\\
\dot B=&4\tilde\varepsilon_1AT+2\tilde\varepsilon_3AS\endaligned\right.$$

Let's consider the case $\varepsilon_3=\varepsilon_2$ (and, hence,
$\tilde\varepsilon_3=\tilde\varepsilon_2$). Then $\Cal L=S-T$ is an
integral of motion. Therefore,
$$\left\{\aligned
\dot Q=&P(2\varepsilon_3\Cal L+(2\varepsilon_3+4\varepsilon_1)T)\\
\dot P=&-QV'(Q^2)(4\varepsilon_3\Cal L+(2\varepsilon_3+4\varepsilon_1)T)
\endaligned\right.\left\{\aligned
\dot B=&A(2\tilde\varepsilon_3\Cal L+(2\tilde\varepsilon_3+
4\tilde\varepsilon_1)T)\\
\dot A=&-BV'(B^2)(4\tilde\varepsilon_3\Cal L+(2\tilde\varepsilon_3+
4\tilde\varepsilon_1)T)
\endaligned\right.$$

Note that the "Hamiltonians" $\Cal H_i=\Cal I^2_i$ are integrals of
motion so put $P=\sqrt{\Cal I_1^2-V(Q^2)}$, $A=\sqrt{\Cal I^2_2-V(B^2)}$
and, hence
$$\left\{\aligned \dot Q=&\sqrt{\Cal I^2_1-V(Q^2)}(4\varepsilon_3\Cal L+
(2\varepsilon_3+4\varepsilon_1)T)\\
\dot B=&\sqrt{\Cal I_2^2-V(B^2)}(4\tilde\varepsilon_3\Cal L+(2\tilde
\varepsilon_3+4\tilde\varepsilon_1)T)\endaligned\right.$$
where $T=\sqrt{\Cal I^2_2-V(B^2)}Q+\sqrt{\Cal I^2_1-V(B^2)}B$. Put
$F_i(x)=\int\frac{dx}{\sqrt{\Cal I^2_i-V(x)}}$, then
$$\left\{\aligned
\dot F_1(Q)=&4\varepsilon_3\Cal L+(2\varepsilon_3+4\varepsilon_1)
(\frac{Q}{F'_2(B)}+\frac{B}{F'_1(Q)})\\
\dot F_2(B)=&4\tilde\varepsilon_3\Cal L+(2\tilde\varepsilon_3+4\tilde
\varepsilon_1)(\frac{Q}{F'_2(B)}+\frac{B}{F'_1(Q)})\endaligned\right.$$

Let's denote $\Theta=F_1(Q)$, $\Xi=F_2(B)$ and put $G_i=F_i^{-1}$,
then $\Xi=\alpha\Theta+\beta t+\gamma$, where $\alpha$, $\beta$
are constants, which may be easily expressed via $\varepsilon_i$,
$\tilde\varepsilon_i$ and $\Cal L$, $\gamma$ is an arbitrary number,
determined by the initial conditions.

Put $G(x,t)=G_1(x)G_2(\alpha x+\beta t+\gamma)$, then $\Theta$ obeys
the following differential equation
$$\dot\Theta=4\varepsilon_3\Cal L+(2\varepsilon_3+3\varepsilon_1)
\frac{\partial G(\Theta,t)}{\partial\Theta}.$$

\subhead 2.3. Representations of the isotopic pairs of noncanonically
coupled oscillators\endsubhead
To quantize the classical dynamics one needs in representations of
algebraic objects underlying it.

\definition{Definition 4 {\rm [5]}}
{\it A representation of the isotopic pair
$(V_1,V_2)$ in the linear space $W$\/} is a pair $(T_1,T_2)$ of mappings
$T_i:V_i\mapsto\End(W)$ such that
$$\align
T_1([X,Y]_A)=&T_1(X)T_2(A)T_1(Y)-T_1(Y)T_2(A)T_1(X),\\
T_2([A,B]_X)=&T_2(A)T_1(X)T_2(B)-T_2(B)T_1(X)T_2(A),
\endalign$$
where $X,Y\in V_1$, $A,B\in V_2$.
A representation of the isotopic pair $(V_1,V_2)$ in
the linear space $W$ is called {\it split\/} iff $W=W_1\oplus W_2$ and
$$\left\{\aligned
(\forall X\in V_1) \left. T_1(X)\right|_{W_2}=0,\ T_1(X):W_1\mapsto W_2,\\
(\forall A\in V_2) \left. T_2(A)\right|_{W_1}=0,\ T_2(A):W_2\mapsto W_1.
\endaligned\right.$$
Otherwords, operators $T(X)$ and $T(A)$ have the form $\left(\matrix 0 & 0
\\ * & 0 \endmatrix\right)$ and $\left(\matrix 0 & * \\ 0 & 0 \endmatrix
\right)$, respectively.
\enddefinition

Not that a split representation of an isotopic
pair $(V_1,V_2)$ defines a representation $T$ of the corresponding anti--Lie
triple system and Lie superalgebra $\frak g(V_1\oplus V_2)$ (or its central
extension $\hat{\frak g}(V_1\oplus V_2)$). The resulted
representation of the Lie superalgebra $\frak g(V_1\oplus V_2)$ always have a
special "polarized" form: $W=W_1\oplus W_2$, $T(V_1):W_1\mapsto W_2$,
$T(V_2):W_2\mapsto W_1$, $\frak g_0(V_1\oplus V_2):W_i\mapsto W_i$.
Note that each representation $(T_1,T_2)$ of the isotopic pair $(V_1,V_2)$ in
the space $W$ defines a split representation $(T^s_1,T^s_2)$ of the same pair
in the space $W_1\oplus W_2$ ($W_i\simeq W$):
$$(\forall X\in V_1)\ T^s_1(X)=\left(\matrix 0 & 0\\ T_1(X) & 0
\endmatrix\right),\quad (\forall A\in V_2)\ T^s_2(A)=\left(\matrix 0 & T_2(A)\\
0 & 0 \endmatrix\right).$$

\subhead 2.4. Quantum dynamics of noncanonically coupled harmonic
oscillators\endsubhead
The formal quantum dynamical equations have
the form
$$\left\{\aligned
\tfrac{d}{dt}\Pn&=-2\varepsilon_1(\Pn\Bn\Q+\Q\Bn\Pn+2\Q\A\Q)-
\varepsilon_3(\R\C\Q+\Q\C\R)\\
\tfrac{d}{dt}\Q&=2\varepsilon_1(\Pn\A\Q+\Q\A\Pn+2\Pn\Bn\Pn)+
\varepsilon_3(\R\C\Pn+\Pn\C\R)\\
\tfrac{d}{dt}\R&=\varepsilon_2(\Pn\A\R+\R\A\Pn-\Q\Bn\R-\R\Bn\Q)
\endaligned\right.$$

$$\left\{\aligned
\tfrac{d}{dt}\A&=-2\tilde\varepsilon_1(\A\Q\Bn+\Bn\Q\A+2\Bn\Pn\Bn)-
\tilde\varepsilon_3(\C\R\Bn+\Bn\R\C)\\
\tfrac{d}{dt}\Bn&=2\tilde\varepsilon_1(\A\Pn\Bn+\Bn\Pn\A+2\A\Q\A)+
\tilde\varepsilon_3(\C\R\A+\A\R\C)\\
\tfrac{d}{dt}\C&=\tilde\varepsilon_2(\A\Pn\C+\C\Pn\A-\Bn\Q\C-\C\Q\Bn)
\endaligned\right.$$

The dynamics is considered in arbitrary representation of the isotopic pair of
noncanonically coupled oscillators. Let's consider such dynamics in the
corresponding split representation. First of all renormalize $c$ and $r$ so
that $R_{p,c}=R_{b,r}$ and $R_{q,c}=R_{a,r}$.

\define\hi{\operatorname{hidden}}
\proclaim{Proposition} Equations of quantum dynamics of
noncanonically
coupled oscillators are a reduction of formal super Heisenberg equations
$$\tfrac{d}{dt}\hat F_t=[\hat H_{\hi},\hat F_t]$$
in $\Cal U(\frak g(V_1\oplus V_2))$ with quadratic quantum Hamiltonian
$$\hat H_{\hi}=\hat R_{q,a}^2+\hat R_{p,b}^2+\hat R_{q,b}^2+\hat R_{p,a}^2+
\hat R_{p,c}^2+\hat R_{q,c}^2$$
\endproclaim

\demo{Proof}The statement of the proposition is verified by
straightforward explicit computation.
\enddemo

So quantum dynamics of noncanonically coupled oscillators admits a hidden
super--Hamiltonian formulation in terms of Lie superalgebra $\frak g(V_1\oplus
V_2)$. It leads to a very important consequence.

\proclaim{Corollary} The quantum dynamics preserves the initial operator
relations:
$$\aligned &\Pn\A\Q-\Q\A\Pn=2\varepsilon_1\Q,\ \Pn\A\R-\R\A\Pn=\varepsilon_2\R,\
\Q\A\R-\R\A\Q=0,\\
&\Pn\Bn\Q-\Q\Bn\Pn=2\varepsilon_1\Pn,\ \Pn\Bn\R-\R\Bn\Pn=0,\
\Q\Bn\R-\R\Bn\Q=-2\varepsilon_2\R,\\
& \Pn\C\Q-\Q\C\Pn=\varepsilon_3\R,\ \Pn\C\R-\R\C\Pn=0,\ \Q\C\R-\R\C\Q=0,\\
&\\
& \A\Pn\Bn-\Bn\Pn\A=2\tilde\varepsilon_1\Bn,\ \A\Pn\C-\C\Pn\A=\tilde\varepsilon_2\C,\
\Bn\Pn\C-\C\Pn\Bn=0,\\
& \A\Q\Bn-\Bn\Q\A=2\tilde\varepsilon_1\A,\ \A\Q\C-\C\Q\A=0,\
\Bn\Q\C-\C\Q\Bn=-\tilde\varepsilon_2\C,\\
& \A\R\Bn-\Bn\R\A=\tilde\varepsilon_3\C,\ \Bn\R\C-\C\R\Bn=0,\ \A\R\C-\C\R\A=0.
\endaligned
$$
\endproclaim

\subhead 2.5. Remark on the quantum dynamics of noncanonically coupled
anharmonic oscillators\endsubhead
It should be marked that the quantization of anharmonic oscillators meets with
a principal difficulty. Namely, each representation of an isotopic pair may be
splitted. After such splitting the elements of $V_1\oplus V_2$ become odd, and
therefore, nilpotent. It provides that {\sl all terms in the classical
equations of motion related to the higher (non--quadratic) terms of a
Hamiltonian are suppressed by the quantization}. Such effect is not realistic.
Of course, one may consider well--defined quantum Hamiltonians by use of the
hidden Lie superalgebraic structure in a way analogous to the proposition above.
The quantum dynamics will preserve the initial operator relations for such
Hamiltonians. However, there is no any {\sl a priori} relation between it and
classical one.

\head\bf 3. Conclusions.\endhead

So the classical and quantum dynamics of noncanonically coupled oscillators
was investigated. It appears that though the classical one is well-defined for
both harmonic and anharmonic oscillators, the quantum one is well-defined for
the noncanonically coupled harmonic oscillators, admits a hidden
(super)Hamiltonian formulation, and thus, preserves the initial operator
relations, but a na\"\i ve quantization of the anharmonic oscillators meets
with principal difficulties.

\head\bf Acknowledgements.\endhead

The author thanks the International Erwin Schr\"odinger Institute for
Mathematical Physics (Vienna, Austria) for a support and a kind hospitality.
This Letter appeared as Preprint ESI no.167 (December 1994) and also as
e--print of LANL e-arch. on solvable and integrable systems
no."solv-int/9503003" (March 1995).

\Refs
\roster
\item"1." Arnol'd V.I., {\it Mathematical methods of classical mechanics},
Springer--Verlag, 1976;\newline
Dubrovin B.A., Novikov S.P., Fomenko A.T., {\it Modern geometry --- methods and
applications}, Springer--Verlag, 1988;\newline
Perelomov A.M., {\it Integrable systems of classical mechanics and Lie
algebras}. Birkhauser--Verlag, 1990;\newline
Karasev M.V., Maslov V.P., {\it Nonlinear Poisson brackets. Geometry and
quantization}. Amer. Math. Soc., R.I., 1993.
\item"2." Juriev D., {\it Russian J. Math. Phys.\/} {\bf 3}:4 (1995) [e-version
(LANL E-Arch. on Solv. Integr. Systems): {\it solv-int/9505001\/} (1995)].
\item"3." Juriev D., {\it Russian J. Math. Phys.}, to appear [e-version
(SISSA E-Arch. on Funct. Anal.): {\it funct-an/9409003\/} (1994)];
Juriev D., e-print (Texas Univ. E-Arch. on Math. Phys.): {\it mp\_arc/95-538\/}
(1995).
\item"4." Faulkner J.R., Ferrar J.C., {\it Commun. Alg.\/} {\bf 8}:11, 993
(1980).
\item"5." Juriev D., {\it Theor. Math. Phys.\/} {\bf 105}:1, 18 (1995) [e-print
version (Texas Univ. E-Arch. on Math. Phys.): {\it mp\_arc/94-267\/} (1994)];
Juriev D., e-print (Duke Univ. E-Arch. on Quantum Alg.): {\it q-alg/9511012\/}
(1995).
\endroster
\endRefs
\enddocument